# Multi-Frequency Resonant Circuit Based Multi-User Emergency Through-the-Earth Communication with Magnetic Induction

Zhenyu Wang, Jianyu Wang, and Wenchi Cheng

State Key Laboratory of Integrated Services Networks Xidian University, Xi'an, China

E-mail:{*23011210705@stu.xidian.edu.cn, wangjianyu@xidian.edu.cn, and wccheng@xidian.edu.cn*}

*Abstract*—Magnetic induction (MI) is an effective technique in emergency through-the-earth communications due to the higher penetration efficiency and lower propagation loss as compared with electromagnetic wave communication. How to cancel the interference between different users and enhance the effectiveness of multi-user transmissions is imperative for the practical application of MI communication. In this paper, we use multi-frequency resonant circuit to establish multiple resonant frequencies for MI communication. The transmissions corresponding to different users operate at different resonant frequencies and multi-user interferences can be naturally mitigated. Numerical results verify our theoretical analyses and show that the proposed system can significantly enhance the performance.

## I. INTRODUCTION

Magnetic induction (MI) is widely utilized in emergency through-the-earth communication due to its higher penetration and lower propagation loss, as compared with electromagnetic wave communication [1], [2]. MI communication finds utility in emergency rescue scenarios, particularly for identifying and locating buried individuals and facilitating basic information transmission. In practical emergency rescue, the above ground node usually needs to communicate with multiple underground nodes. Therefore, to guarantee the efficiency of the post disaster emergency rescue, it is desired to enhance the performance of multi-user emergency MI communication [3].

Multi-user emergency MI communication can be modeled as a single-input multiple-output (SIMO) system [4], where multiple receive antennas are located within the soil medium while the transmit antenna is located in the air. In emergency rescue, the transmit antenna can be suspended by a drone [5]. In multi-user emergency MI communication, a critical challenge that needs to be addressed is the interference among the receive antennas. Due to the magnetic coupling between any two coil antennas, when they sense the magnetic field from the transmit antenna, they will also induce magnetic fields due to the coupling, thereby interfering with other receive coil antennas. In wireless underground sensor networks, interferences can be mitigated to a large extent by controlling the density of sensor placements since sensors are manually deployed. Problems such as network latency can also be effectively controlled [6]. However, in emergency rescue scenarios, the positions of receive antennas are random, and the strength of magnetic coupling among them are unknown, making interference elimination difficult. Hence, in this paper we introduce a multi-frequency resonant circuit to tackle the interference issue.

The concept of multi-frequency resonant circuit was initially proposed in the context of near-field wireless power transfer [6], enabling simultaneous energy and information transmission. By integrating LC resonance circuit into the physical loop of existing coil antennas, additional resonant frequencies can be generated. By controlling the resonant frequencies of the LC resonance circuit, achieved by adjusting the values of inductance or capacitance, the bandwidth allocated at each resonance frequency can be regulated. By constructing multiple resonant frequencies and assigning each resonance frequency to a link, interference issues among receive coil antennas can be addressed to guarantee the stability of multi-user emergency MI communication. As compared with the original single resonant transmission, using a multi frequency resonant circuit can reduce the required SNR by 3dB under the same bit error rate constraint.

The structure of this paper is as follows. In Section II, we establish a multi-user emergency multi-frequency MI communication model and provide a detailed description for the model's parameters and performance evaluation indicators. In Section III, we focus on cross-ground channels. Section IV gives the numerical simulation of the established model. Finally, Section V is a summary of this paper.

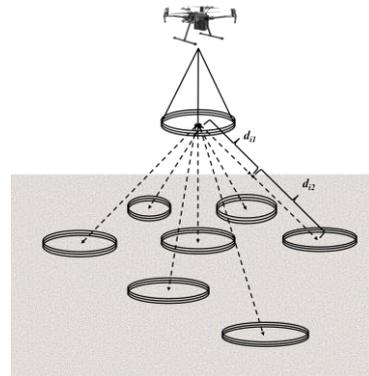

Fig. 1. Multi-user emergency through-the-earth communication with MI.

## II. MULTI-USER EMERGENCY MULTI-FREQUENCY MI COMMUNICATION MODEL

We consider a multi-user emergency MI communication as shown in Fig. 1, where the transmitter is suspended by a drone and transmits signals to multiple underground users. To mitigate the multi-user interferences, we employ multi-frequency resonant circuit to generate multiple resonant frequencies for the coil antenna. The transmitter can allocate different frequency bands to the different links and avoid the multi-user interferences. The equivalent transmit circuit of a traditional single-resonant MI communication system can be equivalently represented as a series LC circuit. To create multiple resonant frequencies in the circuit, a method which adds parallel LC circuit is introduced, as depicted in Fig. 2. It can be observed that the equivalent inductance $L$ of the coil is in series with the compensation circuit composed of $L_i$ and $C_i$, where $i = 1,2,3,\cdots,n$. Let $Z_t$ represent the total impedance of the circuit, then $Z_t$ can be expressed as follows:

$$Z_t = R_t + jX(\omega), \quad (1)$$

where $R_t$ is the coil resistance; $X(\omega)$ is the conductance, with the following expression:

$$X(\omega) = \left[\frac{\omega^2 LC - 1}{\omega C} + \sum_{n=1}^{N} \frac{\omega L_n}{1 - \omega^2 L_n C_n}\right]. \quad (2)$$

Taking the derivative of $X(\omega)$, we can identify the distribution of zeros. By combining with partial fraction expansion, we can obtain the expression of $L_i, i = 1,2,\cdots\cdots,n$, as follows[8]:

$$L_i = \frac{\tilde{X}(\omega)(F_i^2 - \omega^2)}{\omega F_i^2}\bigg|_{\omega=F_i}, \quad (3)$$

$$\tilde{X}(\omega) = L\frac{\prod_{i=0}^{N}(\omega_i^2 - \omega^2)}{-\omega \prod_{i=1}^{N}(F_i^2 - \omega^2)}, \quad (4)$$

where $\omega_i = 2\pi f_i$ is the predetermined resonant angular frequency, $F_i$ can be arbitrarily chosen within the interval $(\omega_{i-1}, \omega_i)$, and $C_i = 1/F_i^2 L_i$. With the above equations, we can calculate the parameters of the multi-frequency resonant circuit according to the required operating angular frequencies.

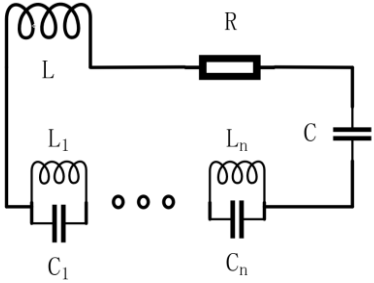

Fig. 2. Multi-frequency resonant circuit.

Then, we derive the path loss between the transmit coil and receive coils. The equivalent circuit between the transmit coil and the $i$-th user is shown in Fig. 3. We use $Z_t$, $Z_{ri}$, and $Z_{Li}$ to represent the transmit impedance, the receive impedance of the $i$-th user, and the load impedance of the $i$-th user, respectively. $Z_{rit} = \omega_i^2 M_i^2/(Z_{ri} + Z_{Li})$ represents the influence of the $i$-th user on the transmitter, $Z_{tri} = \omega_i^2 M_i^2/Z_t$ represents the influence of the transmitter on the $i$-th user. $U_{Mi} = -j\omega_i M_i U_{si}/Z_t$ is the induced voltage at the $i$-th user. Therefore, we can derive the receive power of the $i$-th user and transmit power as

$$P_{ri}(d_{i0}) = \frac{Z_{Li} \cdot U_{Mi}^2}{(Z_{Li} + Z_{tri} + Z_{ri})}, \quad (5)$$

and

$$P_t(d_{i0}) = \frac{U_s^2}{Z_t + Z_{rit}}, \quad (6)$$

respectively, where $d_{i0} = d_{i1} + d_{i2}$ represents the distance from the transmitter to the $i$-th user. Then, we can derive the path loss of the $i$-th user as follows

$$PL_i = \frac{P_{ri}}{P_t} = \frac{\omega_i^2 M_i^2 Z_{Li}(Z_t + \omega_i^2 M_i^2/(Z_{ri} + Z_{Li}))}{(Z_{ri} + Z_{Li} + \omega_i^2 M_i^2/Z_t)^2 Z_t^2}, \quad (7)$$

where $PL_i$ is the path loss from the transmitter to the $i$-th user and $M_i$ is mutual inductance between the transmitter and the $i$-th user. $M_i$ involves a cross-ground channel, we provide a detailed description for it in Section III.

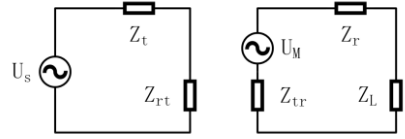

Fig. 3. The equivalent circuit between the transmit coil and the $i$-th user.

Furthermore, the load impedance of the $i$-th user affects the receive power and consequently the path loss. In general, impedance matching can be employed to maximize receive power:

$$Z_{Li} = R_{ri} + \frac{\omega_i^2 M_i^2}{R_t}. \quad (8)$$

In general, when $\omega_i^2 M_i^2/R \ll 1$, we can use the approximation $Z_{Li} \approx R_{ri}$ for impedance matching. The bandwidth for the $i$-th user is the 3dB bandwidth in path loss.

### III. CROSS-GROUND CHANNEL

As shown in Fig. 1, the transmit coil antenna generates magnetic fields in the air, which radiate in various directions. The downward radiating magnetic field will pass through the air ground interface and radiate into the soil medium. In this section, we pay attention to the propagation model of the cross-ground magnetic field radiation.

In fact, as early as the last century, the propagation of magnetic dipoles in layered media has been studied [10], [10]. The results obtained are accurate, but for practical applications, the model complexity appears to be high. Since the coil antenna can radiate and receive very little energy in the far field, we use the approximate near-filed model proposed in [11] in the following for calculation simplicity. We use $a$, $N$, and $I$ to denote the radius, number of turns, and current of the loop, with subscript $t$ or $r$ indicating whether they belong to the transmit or receive loop. Additionally, we use $\mu$, $\epsilon$, and $\sigma$ to represent the permeability, permittivity, and conductivity, respectively. Subscript 1 represents the parameters in medium 1 (air), and subscript 2 represents the parameters in medium 2 (soil). The transmission propagation constant corresponding to the $i$-th frequency in medium 1 and medium 2 are denoted by $k_{1,i} = \omega_i\sqrt{\mu_1 \epsilon_1}$ and $k_{2,i} = \omega_i\sqrt{\mu_2 \epsilon_2}$, respectively.

The magnetic field corresponding to the $i$-th frequency radiated by the transmitter in space can be expressed as [12]:

$$\boldsymbol{h}_i(k_{1,i}d) = \boldsymbol{h}_{fi}(k_{1,i}d) + \boldsymbol{h}_{rni}(k_{1,i}d) + \boldsymbol{h}_{ni}(k_{1,i}d), \quad (9)$$

where $\boldsymbol{h}_{fi}$ represents the far field, $\boldsymbol{h}_{rni}$ denotes the radiation near field, $\boldsymbol{h}_{ni}$ represents the near field, and $d$ is the distance. Their expressions are as follows:

$$\boldsymbol{h}_{fi}(k_{1,i}d) = [0, -k_{1,i}^2 \xi_y/d, 0]^T; \quad (10)$$

$$\boldsymbol{h}_{rni}(k_{1,i}d) = [jk_{1,i}\xi_x/d^2, jk_{1,i}\xi_y/d^2, 0]^T; \quad (11)$$

$$\boldsymbol{h}_{ni}(k_{1,i}d) = [\xi_x/d^3, \xi_y/d^3, 0]^T, \quad (12)$$

where $\xi_x = a_t^2 N_t I_t \cos\theta e^{jk_{1,i}d}/2$ and $\xi_y = a_t^2 N_t I_t \sin\theta e^{jk_{1,i}d}/4$. For the near-field propagation in MI communication that we focus, it is primarily influenced by $\boldsymbol{h}_{ni}$ and the first term of $\boldsymbol{h}_{rni}$, if we denote $\boldsymbol{h}_{rni}^{\vec{r}}(k_i d) = [jk_i\xi_x/d^2, 0, 0]^T$, we can obtain the following approximate result for the $i$-th user [11]:

$$\begin{aligned}\boldsymbol{h}_i^s \approx &\left([\boldsymbol{h}_{ni}(k_{1,i}d_{i1}) + \boldsymbol{h}_{rni}^{\vec{r}}(k_{1,i}d_{i1})]\right.\\&\left.\oslash [\boldsymbol{h}_{ni}(k_{2,i}d_{i1}) + \boldsymbol{h}_{rni}^{\vec{r}}(k_{2,i}d_{i1})]\right)\\&\odot [\boldsymbol{h}_{ni}(k_{2,i}d_{io}) + \boldsymbol{h}_{rni}^{\vec{r}}(k_{2,i}d_{io})],\end{aligned} \quad (13)$$

where $\oslash$ and $\odot$ represent Hadamard division and production respectively, $d_{i1}$ and $d_{i2}$ are shown in Fig. 1, $d_{i0} = d_{i1} + d_{i2}$, $\boldsymbol{h}_i^s$ represents the magnetic field around the $i$-th user, and the superscript $s$ represents the spherical coordinate system.

We aim to calculate the mutual inductance between the transmitter and the $i$-th user from $\boldsymbol{h}_i^s$. Therefore, we need to know $\boldsymbol{h}_i^s$ in the z-direction. However (13) is derived in spherical coordinates, thus coordinate transformation is required:

$$\mathbf{R} = \begin{pmatrix} \sin\theta\cos\phi & \sin\theta\sin\phi & \cos\theta \\ \cos\theta\cos\phi & \cos\theta\sin\phi & -\sin\theta \\ -\sin\phi & \cos\phi & 0 \end{pmatrix}. \quad (14)$$

Then the magnetic fields expressed in the cartesian coordinate system is

$$\boldsymbol{h}_i^c = \boldsymbol{h}_i^s \mathbf{R}^T. \quad (15)$$

Let $\boldsymbol{h}_i^z$ denote the expression for the magnetic field in the z-direction. After computation, we can obtain

$$\boldsymbol{h}_i^z = BD[A\cos\theta^2 - \sin\theta^2/2]/2, \quad (16)$$

where $\theta$ represents the angle between the line connecting the center points of the $i$-th user and the transmit coil and the vertical direction,

$$A = \frac{(1 + jk_{1,i}d_{i1})(1 + jk_{2,i}d_{i0})}{1 + jk_{2,i}d_{i1}}, \quad (17)$$

$$B = a_t^2 N_t I_t e^{j(k_{1,i}d_{i1} + k_{2,i}d_{i0} - k_{2,i}d_{i1})}, \quad (18)$$

and

$$D = \frac{d_{i2}^3}{d_{i1}^3 d_{i0}^3}. \quad (19)$$

This model can be further simplified, for instance, when $k_{2,i}d_{i2} \ll 1$, we can assume $e^{jk_{2,i}d_{i2}} = 1$, which leads to the $\boldsymbol{h}_i^z \approx \boldsymbol{h}_{ni}\mathbf{R}^T$. The approximation is also widely used [14], [14]. With $\boldsymbol{h}_i^z$, we can calculate mutual inductance based on the magnetic flux model as follows:

$$M_i = \frac{\phi_i}{I}, \quad (20)$$

$$\phi_i = \int_s \mu_2 N_{ri} \boldsymbol{u}_i^t \cdot \boldsymbol{h}^r ds, \quad (21)$$

where $\boldsymbol{u}_i^t$ is the unit vector in the axial direction of the coil of the $i$-th user, according to our previous derivation, we have $h_z^u = \boldsymbol{u}_i^t \cdot \boldsymbol{h}^r$.

In addition, as soil is a loss medium, the medium around the $i$-th user absorbs a portion of the magnetic field. To characterize this phenomenon, the antenna inductance of the $i$-th user will be a complex number, with the actual part representing inductance and the imaginary part representing near-field loss. We still consider modeling inductance using magnetic flux, with an approximate expression as follows:

$$l_i \approx \frac{\mu_2 \pi a_i^2 N_i \xi_x}{a_i^3} = \frac{1}{2}\mu_2 \pi a_i N_i^2 e^{jk_{2,i}a_i}. \quad (22)$$

Environmental factors can also influence a series of parameters, thereby affecting path loss. In the cross-ground channel model, changes in soil moisture content profoundly affect the dielectric constant and propagation constant of the soil medium, thus impacting the self-inductance of the receive coil and the mutual inductance between the receive coil and the transmit coil, ultimately affecting the system's path loss [15][15]. However, the model established in this paper is aimed at emergency rescue scenarios, where time for rescue operations after a disaster is very limited. Therefore, it can be assumed that environmental factors remain stable during the rescue time and do not affect the system performance.

IV. NUMERICAL SIMULATIONS AND NUMERICAL ANALYSIS

In this section, we employ numerical evaluation to assess the performance of the established model, with performance metrics including path loss, bandwidth, and channel capacity. We consider medium 1 as air and medium 2 as soil, with the dielectric constant of soil being 5.343 and the conductivity being $7.680 \times 10^{-8} \, S/m$. The transmit coil is placed in medium 1, while the receive coil is positioned in medium 2.

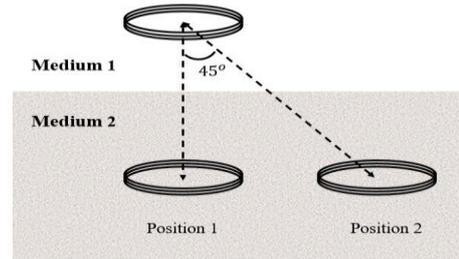

Fig. 4. Schematic diagram of the simulation.

The receive coil has two configurations: one aligned parallel to the transmit coil, and the other offset by 45 degrees, as depicted in Fig. 4. In the simulation, the transmit coil has a radius of 1m and 10 turns, wound with copper wire, while the receive coil has a radius of 0.15m and 10 turns, also wound with copper wire. The vertical distance between the transmit and receive coils is 10m, with the vertical distance from the transmit coil to the soil being 2m, and from the receive coil to the soil surface being 8m. The transmit coil utilizes a multi-frequency resonant circuit, with resonant frequencies of 1MHz and 5MHz.

Comparing Figs. 5 and 6, we can see that when both the transmit and receive coils are equipped with multi-frequency resonant circuit, the path loss significantly decreases at the resonant frequencies of 1MHz and 5MHz, reaching a minimum of around 18dB, which is sufficient for communication requirements. As the transmit circuit must use a multi-resonant circuit to achieve interference-free multi-user communication, we propose that the receive end only needs a single resonant circuit to meet communication needs. Fig. 6 illustrates that when the receiver only requires a single resonant circuit, the path loss curve at 5MHz becomes smoother, indicating an expanded bandwidth at 5MHz. In terms of bandwidth values, when the receive end also employs a multi-frequency resonant circuit, the 3dB bandwidth at 5MHz is approximately 103kHz. In contrast, when the receive end uses a single resonant circuit, the 3dB bandwidth at 5MHz increases to approximately 132kHz, a substantial 28% improvement in bandwidth. Comparing Figs. 7 and 8, the 3dB bandwidth at 5MHz increases from 159kHz to 167kHz. Therefore, for the proposed model, a multi-frequency resonant circuit should be used at the transmit end, while the receive end should employ a single resonant circuit corresponding to the communication frequency. This approach maximizes communication speed.

Comparing Figs. 5 and 7, it's evident that when the receive coil is offset by 45 degrees, there's a noticeable increase in path loss. When vertically positioned, the minimum path loss is around 18dB, whereas when offset by 45 degrees, it's around 44dB, resulting in a difference of approximately 27dB. This indicates a substantial attenuation rate of magnetic induction communication with changes in distance and angle, imposing significant limitations on such communication in emergency rescue scenarios. Comparing the bandwidth at 5MHz between Fig. 5 and Fig. 7, the bandwidth in Fig. 7 is approximately 159kHz, even larger than that in Fig. 6 at 5MHz. We attribute this to the increase in distance and changes in angle, which lead to a higher increase in losses at the resonant frequency compared to losses at non-resonant frequencies, resulting in a smoother curve at the resonant frequency.

We can know from Fig. 9 that it is evident that using a single resonant circuit significantly enhances the channel capacity compared to using multiple resonant circuit. Moreover, as the signal-to-noise ratio increases, the proportion of enhancement grows, approximately doubling the original channel capacity. This enhancement ratio surpasses that of the bandwidth increase because employing a single resonant circuit results in a smoother path loss curve. Although it doesn't decrease the loss at the resonance frequency, the smoother curve reduces the loss corresponding to the resonance frequency within the bandwidth, leading to an increase in channel capacity. Therefore, the proportion of channel capacity enhancement using a single resonant circuit exceeds that of the bandwidth increase. In Fig. 9, Env5 represents the receive coil offset by 45 degrees and using a single resonant circuit, while Env6 represents the receive coil offset by 45 degrees and using multiple resonant circuit. Comparing Env5 and Env6, the enhancement using a single resonant circuit is minimal in the 45-degree offset scenario, unlike the non-offset scenario. Additionally, the 45-degree offset increases path loss, leading to lower channel capacity compared to the non-offset scenario. However, we observe that when the receive coil is offset by 45 degrees, the growth rate of channel capacity is higher compared to the non-offset scenario. According to the earlier bandwidth analysis, this is due to the fact that in the 45-degree offset scenario, the bandwidth at the resonant frequency is higher than it was prior to the offset. As a result, as the signal-to-noise ratio increases, the growth rate of channel capacity in the 45-degree offset scenario surpasses that of the non-offset scenario.

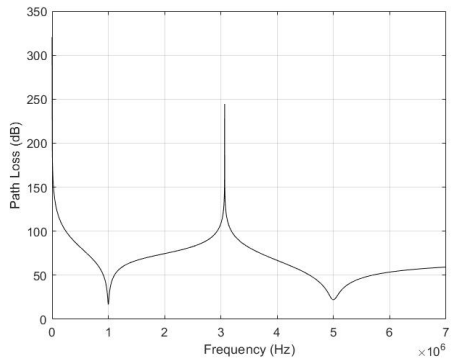

Fig. 5. Path loss using multi-frequency resonant circuit for both transmit and receive coils, with the receive coil located at position 1.

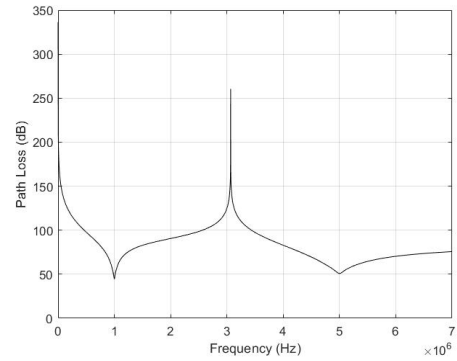

Fig. 7. Path loss using multi-frequency resonant circuit for both transmit and receive coils, with the receive coil located at position 2.

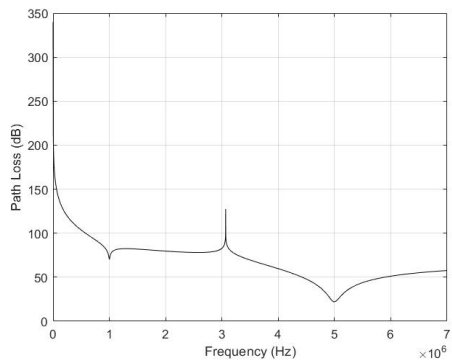

Fig. 6. Path loss using multi-frequency resonant circuit for transmit coil and single resonant circuit for the receiver coil operating at 5MHz, located at position 1.

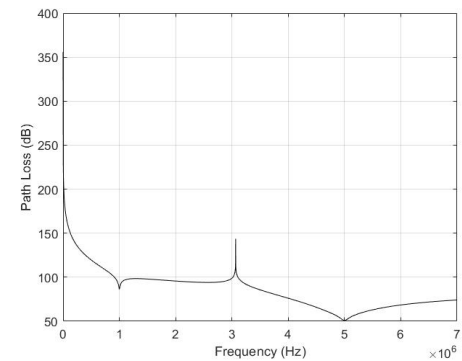

Fig. 8. Path loss using multi-frequency resonant circuit for transmit coil and single resonant circuit for the receiver coil operating at 5MHz, located at position 2.

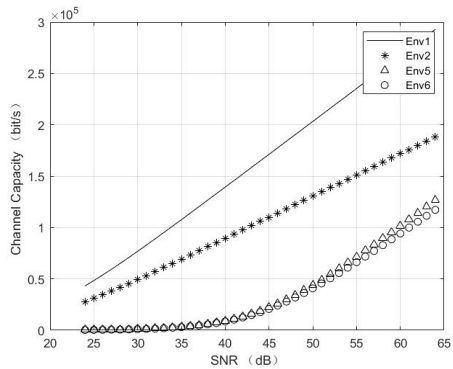

Fig. 9. Comparison of channel capacity in four scenarios, where the transmitters use a multi frequency resonant circuit. Env1 and Env5 are receivers using a multi frequency resonant circuit, Env2 and Env6 are receivers using a single resonant circuit, where the receivers of Env1 and Env2 are located at position 1 and the receivers of Env5 and Env6 are located at position 2.

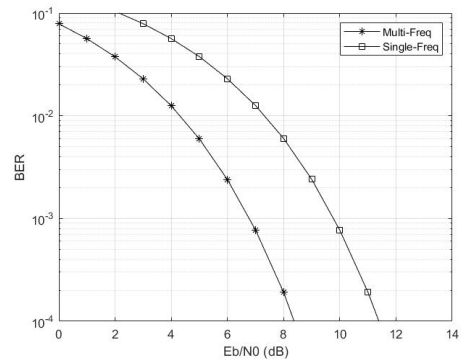

Fig. 10. Comparison of error rates using multi-frequency resonant circuit and single resonant circuit for transmit.

In Fig. 10 we compare the error rate of single resonance multi-user emergency communication with that of multi resonance multi-user emergency communication. We can observe that under the same bit error rate limit, the system performance of the transmitter using a single resonant circuit is 3dB worse than that using a multi frequency resonant circuit. This is because when the transmitting antenna uses a multi frequency resonant circuit to communicate with multiple users, different frequency bands can be allocated to each link to avoid interference, and the transmitting power can be used entirely to transmit information to the target user. However, when using a single resonant circuit to communicate with two users at the same time, only one frequency band can be used to transmit information, and the information of the two users will be stacked together. When the transmitting power is constant, it is equivalent to only half of the power being used to send information to the target user. Therefore, the performance will be 3dB worse under the same bit error rate.

## V. CONCLUSION

In this paper we proposed a multi-user emergency multi-frequency communication model for emergency rescue and addressed the inherent interference issues in MI communication. We introduced a novel physical structure, that is, multi frequency resonant circuit to generate multiple resonant frequencies, for the coil antenna. The allocation of different frequency bands to each link in multi-user emergency communication effectively reduces the multi-user interferences. Performance indicators including path loss, bandwidth, channel capacity, and bit error rate of the proposed scheme was evaluated with simulations, which verifies the effectiveness of the proposed model.